\documentclass{rstransa}

\usepackage{graphicx}
\usepackage{dcolumn}
\usepackage{bm}
\usepackage{hyperref}
\usepackage{textgreek}
\usepackage{xcolor}
\usepackage[caption=false]{subfig}

\begin{document}

\title{Theory and Application of Cavity Solitons in Photonic Devices }

\author{Gian-Luca Oppo and William J. Firth}
\address{SUPA and Department of Physics, University of Strathclyde, 107 Rottenrow,
Glasgow, G4 0NG, Scotland, UK}

\subject{Nonlinear and Quantum Optics}
\corres{g.l.oppo@strath.ac.uk}

\begin{abstract}
Driven optical cavities containing a nonlinear medium support stable dissipative solitons, cavity solitons, in the form of bright or dark spots of light on a uniformly-lit background. Broadening effects due to diffraction or group velocity dispersion are balanced by the nonlinear interaction with the medium while cavity losses balance the input energy. The history, properties, physical interpretation and wide application of cavity solitons are reviewed. Cavity solitons in the plane perpendicular to light propagation find application in optical information processing, while cavity solitons in the longitudinal direction produce high quality frequency combs with applications in optical communications, frequency standards, optical clocks, future GPS, astronomy and quantum technologies.
\end{abstract}

%

\maketitle


 
\section{Introduction}\label{sec:intro}
Rodney Loudon's work has been pivotal for the establishment and success of research in quantum optics, quantum electronics and photonics in the UK. Quantum electronics deals with the interaction of radiation with discrete energy levels in a medium (as in a maser or a laser). Rodney's original work in the early 1960's was indeed in quantum electronics about the Raman effect in crystals \cite{Loudon1964}. Nonlinear optics bridges themes of quantum optics and quantum electronics to describe the behaviour of light in media where the polarization density responds non-linearly to the electric field of light. In his seminal book 'The Quantum Theory of Light' \cite{RodneyBook2000}, Rodney beautifully describes single-mode laser theory, the dynamics of atom-radiation systems, resonance fluorescence and nonlinear quantum optics. Rodney's work opened the doors for the flourishing of these research areas in the UK. Cavity Solitons discovered at Strathclyde in the 1990s and reviewed here, is a story of a great success for nonlinear optics, quantum electronics and photonics in the UK.

As well as a great scientist, Rodney was a real gentleman. Never crossed, always polite, a pleasure to discuss and learn about topics in quantum and nonlinear optics. Rodney had been a visiting professor at Strathclyde for many years and is sadly missed by us all. This paper is dedicated to his work and memory.  

Optical solitons are pulses (or beams) of light in which light-matter nonlinearity counter-balances dispersion or diffraction, leading to a robust structure which propagates without change of form. Strictly speaking, true solitons are exact solutions of integrable nonlinear partial differential equations \cite{Dauxois2010}, but in nonlinear optics mathematical nicety is often foregone in favour of physical robustness. Thus the term ‘optical soliton’ is now routinely used for any pulse or beam of light in which dispersion and/or diffraction is compensated on the average by nonlinearity. Many interesting solitons and soliton structures have been predicted and observed in nonlinear optical systems. Here we focus on a class of dissipative solitons, localised bright or dark spots in driven optical cavities, the cavity solitons (CSs). They share some properties with conservative propagation solitons which have been developed as ‘bits’ for long-haul fibre-optic communications \cite{Kivshar2003,Agrawal2006}. Such structures can also be natural ‘bits’ for parallel processing of optical information, especially if they exist in semiconductor micro-resonators. In the longitudinal direction (along the cavity axis), CSs represent optimal pulses with pyramidally shaped frequency spectra. These spectra can easily span more than one octave in the frequency (or wavelength) domain while being formed by thousand of components separated by the free spectral range of the optical cavity round trip time. These kinds of spectra are known as frequency combs and CS generated frequency combs have taken the optics community by storm. CS generated frequency combs have found applications in frequency standards, optical clocks, optical communications, future GPS, astronomy. and quantum technologies.
 
There are several review articles in the literature about CSs, dissipative solitons \cite{Firth1998,Ackemann2005,Ackemann2009} and temporal CSs \cite{Coen2016}. Here, while reviewing the subject of CSs in photonic devices, we connect the early literature of the 1990s with the most recent developments and applications of temporal CSs since the objects of interest, the CSs, are exactly the same from the mathematical point of view. The paper is organised as follows. In the next section we introduce and derive the most common models for the study of CSs, the Lugiato-Lefever equation (LLE) for the Kerr case and the saturable absorber equation for the full optical nonlinearity. We then describe in Section 3 the instability of the homogeneous stationary solutions to spatial wave vectors leading to the formation of CSs in both the LLE and saturable absorber models. Section 4 is dedicated to historical highlights and properties of CSs in both cases of diffraction (transverse plane) and group velocity dispersion (longitudinal direction). Section 5 is dedicated to the applications of CSs in optical memories, delay lines, optical communications, information processing and frequency combs. Section 6 contains the conclusions and outlook.

\section{Model Equations}
\begin{figure}
	\includegraphics[width=0.50\columnwidth]{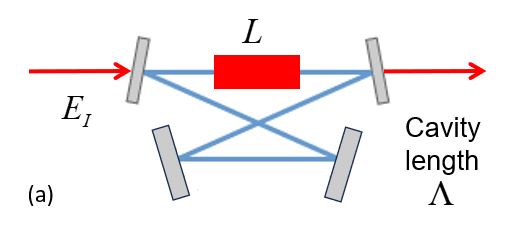}
	\includegraphics[width=0.49\columnwidth]{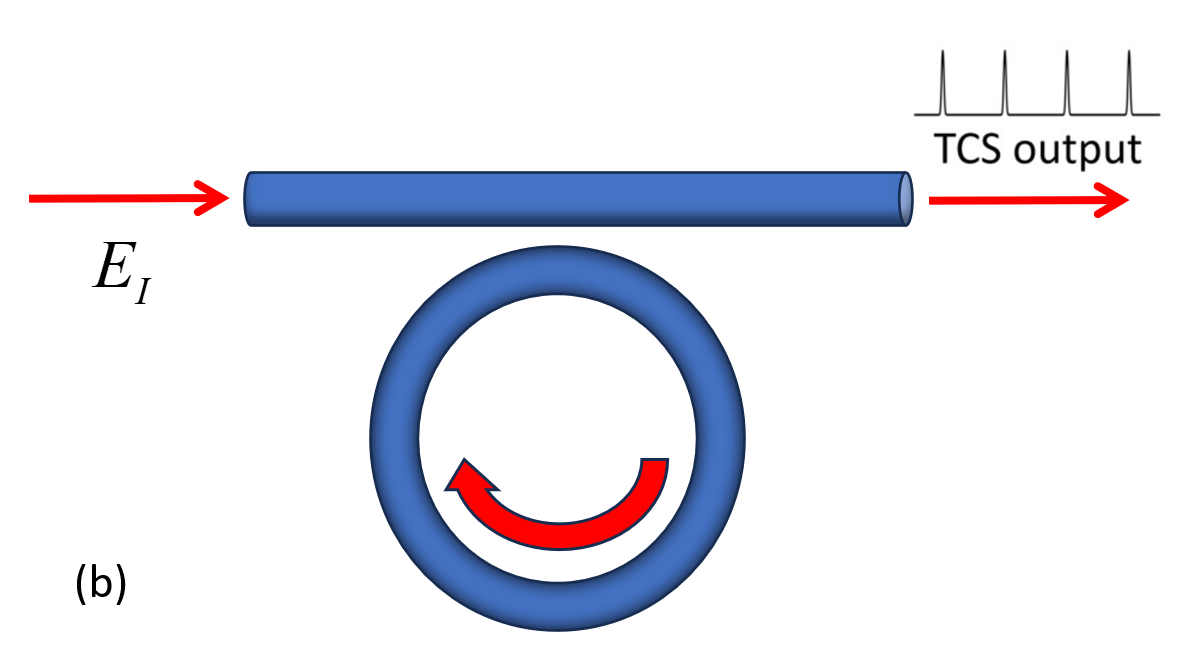}
	\caption{(a) Bow tie cavity for the case with diffraction. (b) Ring cavity for the case with group velocity dispersion with a typical output formed by a train of Temporal Cavity Solitons (TCSs).}
	\label{Cavities}
\end{figure}

We briefly derive the standard models for the generation and observation of CSs. We consider the optical cavities shown in Fig.~\ref{Cavities} where an input light beam $E_I$ leads to a resonated electrical field that interacts with a nonlinear optical medium. In particular we use the bow tie cavity configuration of Fig.~\ref{Cavities}(a) for the diffractive case (although the physics of more practical Fabry-Perot cavities is very similar \cite{Firth2021}) where the transverse plane perpendicular to the direction of propagation $z$ is identified by the coordinates $(x,y)$. We use instead the ring resonator configuration of Fig.~\ref{Cavities}(b) for the dispersive case where light propagation is guided in the transverse direction.

\subsection{The Lugiato-Lefever Equation}
We start from the standard propagation equation of coherent light $E$ in a Kerr medium, the Nonlinear Schr\"odinger Equation (NLSE)~\cite{NLSE2005}:
\begin{equation}
\partial_z E + \frac {n}{c} \,\,\, \partial_t E = \frac{i}{2k} \nabla^2 E + i \eta |E|^2 \, E
\label{NLSE}
\end{equation}
where $\partial$ denotes partial derivatives with respect to the given variable, $z$ is the propagation coordinate, $n$ is the linear refractive index of the medium, $c$ is the speed of light in vacuum, $t$ is time, $k$ is the light wave vector, and $\eta$ is equal to $\pm 1$ for focusing or defocusing media respectively. The term $\nabla^2 E$ with $\nabla^2 = \partial^2_x+\partial^2_y$ describes diffraction, where $(x,y)$ are the coordinates of the plane perpendicular to the optical axes. The term $(2k)^{-1} \nabla^2 E $ can be replaced by $\gamma \partial^2_t E$ to account for group velocity dispersion with the constant $\gamma$ being positive in the anomalous dispersion regime and negative in the normal dispersion regime. 

In the bow tie cavity of length $\Lambda$ with a crystal of length $L$, Fig.~\ref{Cavities}(a), at $z=0$ (entrance of the crystal) the boundary condition is
\begin{eqnarray}
E(x,y,0,t) &=& e^{D} \, E \left(x,y,L,t-\frac{\Lambda - L}{c} \right ) + 
\sqrt{T} E_{I}(x,y) \label{BC} \\
D &=& {\mathrm ln} \, \sqrt{R} - i \Theta + (\Lambda - L) 
\frac {i} {2k} \nabla^2 \\
\Theta &=& \frac {\omega_c - \omega} {c} \Lambda 
\;\;\;\;\;\;\;\;\;\;\;\; T = 1 - R ,
\end{eqnarray}
where $R (T)$ are the mirror reflectivities (transmittivities) for the field intensity, $E_{I}$ is the c.w. input field at frequency $\omega$, while $\omega_c$ is the frequency of the longitudinal cavity mode closest to $\omega$. 
In the following we use the transmittivities of the mirrors as small parameters, i.e.
\begin{equation}
\sqrt{T} = \varepsilon \ll 1 \, .
\end{equation}
Using this condition the standard, but detailed, steps presented in the Appendix enable the problem to be reduced to a single evolution equation. We can introduce renormalised parameters  
\begin{eqnarray}
\label{NPAR2}
\theta = \frac{\Theta} {T/2}; \;\;\;\;\;\;\;\;
a = \frac {\Lambda-L} {kT}; \;\;\;\;\;\;\;\; \kappa = \frac{T}{2\tau}
\end{eqnarray}
to obtain:
\begin{eqnarray}
\partial_{\kappa t'} E = E_{I} -(1+i\theta) E + i \eta |E|^2 E + i a \nabla^2 E . 
\label{LLE}
\end{eqnarray}
where $E_I$ has been normalised by $\sqrt{T/2}$. Eq. (\ref{LLE}) is the renowned spatial Lugiato-Lefever LLE model \cite{Lugiato1987}.

As noted in the Appendix, closely analogous steps can be taken in the case of a fibre ring cavities and solid-state microresonators, see Fig.~\ref{Cavities}(b), in the presence of group velocity dispersion \cite{Haelterman1992,Castelli2017} to obtain a temporal LLE given by:
\begin{eqnarray}
\partial_{\kappa t'} E = E_{I} -(1+i\theta) E + i \eta |E|^2 E + i \beta \partial^2_\tau E . 
\label{FTLLE}
\end{eqnarray}
where $\beta$ is the group-velocity dispersion coefficient, $\tau$ is the fast time during a round trip of the cavity while $\kappa t'$ is known as the slow time since it describes how the intra-cavity field evolves over many round trips \cite{Coen2016}. An interesting feature of the temporal LLE is that the coefficient $\beta$ can be either positive or negative in the anomalous or normal dispersion regimes, respectively. 

An important consideration is that, while the phenomena of diffraction and group velocity dispersion have different physical origins, Eqs.~(\ref{LLE}) and (\ref{FTLLE}) are mathematically identical. This means that, for example, in one spatial dimension, homogeneous and localized solutions, their instabilities and their dynamics found in one of these equations have immediate counterparts in the other for the same parameter values. Neglecting this trivial fact can only lead to confusion, omissions and unnecessary repetitions~\cite{Kippenberg2018}. A historical review of the LLE is provided in \cite{Lugiato2018}.

\subsection{The Saturable Absorber Equation}
If we consider propagation in a saturable absorber instead of a pure Kerr medium, the NLSE changes into:
\begin{equation}
\partial_z E + \frac {n}{c} \,\,\, \partial_t E = \frac{i}{2k} \nabla^2 E - \frac{Q (1-i\Delta) E}{1+\Delta^2+|E|^2}
\label{SA}
\end{equation}
where $Q$ is a numerical factor proportional to the atomic density in the medium and $\Delta$ is proportional to the detuning between the input laser frequency and the atomic frequency of a two energy level system. Note that the intra-cavity intensity has been normalised by the saturation intensity. 
By repeating the MFL steps outlined in the previous section and in the Appendix it is possible to obtain:
\begin{eqnarray}
\partial_{\kappa t'} E = E_{I} -(1+i\theta) E - \frac{2C (1-i\Delta) E}{1+\Delta^2+|E|^2} + i a \nabla^2 E ,
\label{CavitySA}
\end{eqnarray}
where the constant $C$ is known as the cooperativity factor and is proportional to both the square of the matrix element of the dipole transition and the atomic density. It is easy to see that in the limit of large $|\Delta|$ and small intra-cavity intensities $|E|^2$, Eq.~(\ref{CavitySA}) reduces to the LLE~(\ref{LLE}). Another interesting limit is the atomic resonance case of $\Delta=0$ where Eq.~(\ref{CavitySA}) becomes:
\begin{eqnarray}
\partial_{\kappa t'} E = E_{I} -(1+i\theta) E - \frac{2C \, E}{1+|E|^2} + i a \nabla^2 E . 
\label{CavityPureA}
\end{eqnarray}
This is known as the purely absorptive case.

\section{Plane Wave Instability and Cavity Solitons}
The CSs described here typically occur when stationary, i.e. $\partial_{\kappa t'} E =0$, homogeneous, i.e. $\nabla^2 E=0$, solutions coexist with spatially modulated structures (patterns). The bifurcations where homogeneous stationary states are unstable to spatial wave vectors $K$ are known as Turing instabilities \cite{Lugiato1987}. CSs exist where a localised perturbation does not spread transversely, enabling a spatially localised stationary state. Since there is no transverse spreading, other localised states can be created nearby, but remain independent, forming an array of independent "bits". Thus an array of $n$ sites can support $2^n$ different states, and thus has a huge information capacity. For the bistability of Turing patterns and homogeneous stationary states to exist, the instability of the homogeneous states has to be subcritical when increasing the input amplitude $E_I$ while keeping the detuning $\theta$ fixed \cite{Harkness2002,McSloy2002}. Hence, the "cavity soliton" region is in general found for input amplitudes $E_I$ below the Turing instability threshold of the homogeneous stationary states.

\subsection{Cavity Solitons in Kerr Media}
\begin{figure}
	\includegraphics[width=0.32\columnwidth]{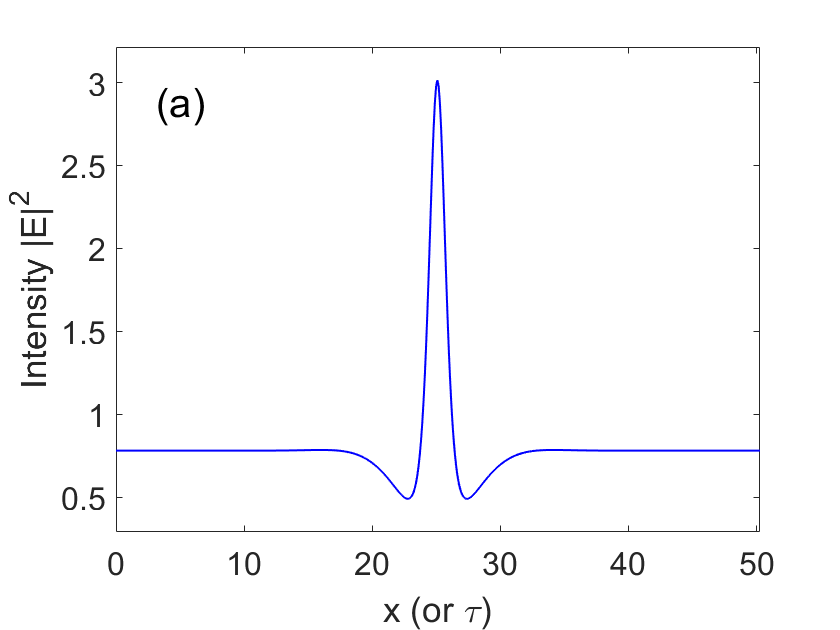}
	\includegraphics[width=0.32\columnwidth]{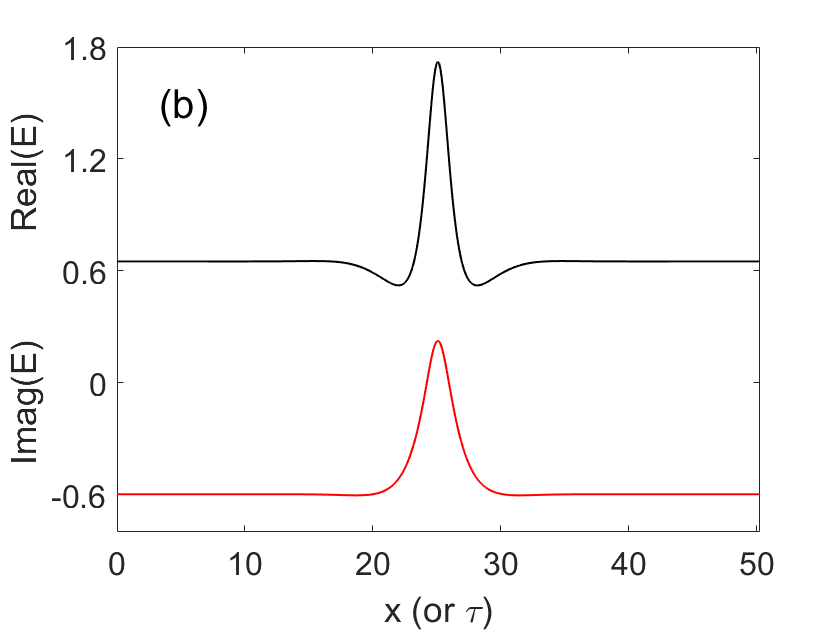}
	\includegraphics[width=0.32\columnwidth]{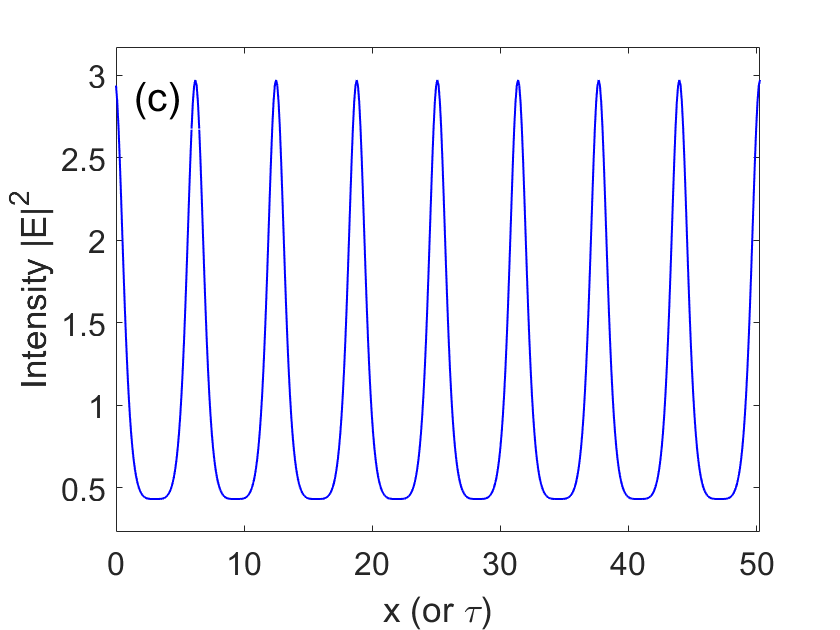}
	\caption{(a) Intensity of a CS of the LLE model (\ref{LLE}) for $E_I=1.2$ and $\theta=1.7$. (b) Real and imaginary parts of the CS field. (c) Stable pattern at the same parameter values.}
	\label{CS}
\end{figure}

For self-focusing Kerr nonlinearities ($\eta=1$), Eqs.~(\ref{LLE}) and (\ref{FTLLE}) admit homogeneous stationary solutions $E_s$ obeying the implicit equation
\begin{equation}
E_I^2=|E_s|^2 \left[ 1 + \left( \theta- |E_s|^2 \right)^2 \right] .
\label{HSS}
\end{equation} 
The steady-state curve of $|E_s|^2$ as a function of $E_I^2$ is single-valued for $\theta < \sqrt{3}$ and 
S-shaped with possible optical bistability, for $\theta > \sqrt{3}$. We introduce perturbations proportional to $\exp(\lambda \kappa t') \exp(i \vec{K} \cdot \vec{x})$ where $\vec{x}=(x,y)$ and $\vec{K}=(K_x,K_y)$ in two dimensions, to perform the linear stability analysis of the homogeneous stationary solutions. Following \cite{Scroggie1994}, one finds that these solutions are unstable to the growth of modulations in the wave vector interval of
\begin{equation}
\eta \left( 2 |E_s|^2 - \theta \right) - \sqrt{ |E_s|^4 - 1 } \;\;\;\; < aK^2 < \;\;\;\; 
\eta \left( 2 |E_s|^2 - \theta \right) + \sqrt{ |E_s|^4 - 1 }
\end{equation}
When plotting these curves in a $(aK^2, |E_s|^2)$ diagram one can find that given the input amplitude $E_I$ and the detuning $\theta$ there are critical values $(aK_c^2=2-\theta, |E_s|_c^2=1)$ corresponding to minima of these curves for $\theta<2$. For $\sqrt{3} < \theta < 2$ the entire upper branch of the hysteresis cycle of the homogeneous stationary solutions is unstable to patterns as well as a segment of the lower branch, whereas for $\theta> 2$ the upper branch is still unstable but the lower branch is stable. Moreover, the Turing instability leading to patterns is supercritical for $\theta<41/30$ and subcritical for $\theta>41/30=1.3666..$ \cite{Lugiato1987}. 
The subcritical condition is ideal to obtain simultaneously stable patterns and homogeneous stationary states. Following \cite{Scroggie1994}, we select $E_I=1.2$ and $\theta=1.7$ where there is no bistability of homogeneous states but stable homogeneous solutions can coexist with a stable branch of periodically modulated patterns. By starting from a homogeneous input beam $E_I$ with a strong perturbation in its middle, we numerically simulate the LLE model with a split-step Fourier method \cite{Harkness2002,McSloy2002} and observe the formation of a CS. After a short transient, the perturbation is removed and the flat input beam restored but the CS solution survives indefinitely as shown in Fig.~\ref{CS}. The CS balances the focusing nature of the Kerr nonlinearity with the diffractive/dispersive broadening while the cavity losses are balanced by the input power \cite{Scroggie1994,Firth1998}. The CS intensity peak is shown Fig.~\ref{CS}(a), its real and imaginary parts in Fig.~\ref{CS}(b) and a coexisting pattern solution in Fig.~\ref{CS}(c). Note that the plot of the real part of $E$ faithfully and accurately reproduces Fig. 11(c) of \cite{Scroggie1994} in spite of the enormous progress made by computers in the last thirty years. As described below, Fig. 11(c) of \cite{Scroggie1994} is the first ever evidence of a CS in the LLE model and was discovered at the University of Strathclyde.

It is important to note that the horizontal scale of Fig.~\ref{CS} can either be the transverse coordinate $x$ with the LLE in the presence of diffraction or the fast time $\tau$ of the LLE in the presence of anomalous dispersion. The two solutions are of course identical and for this reason a large part of the theoretical/numerical work done from the early 1990s in the diffractive case has an immediate application in the group velocity dispersion case. Unless one uses intra-cavity telescopes to drastically modify the diffraction length, the coefficient $a$ in front of the Laplacian operator in Eq.~(\ref{LLE}) is always positive. 

\begin{figure}
	\includegraphics[width=0.32\columnwidth]{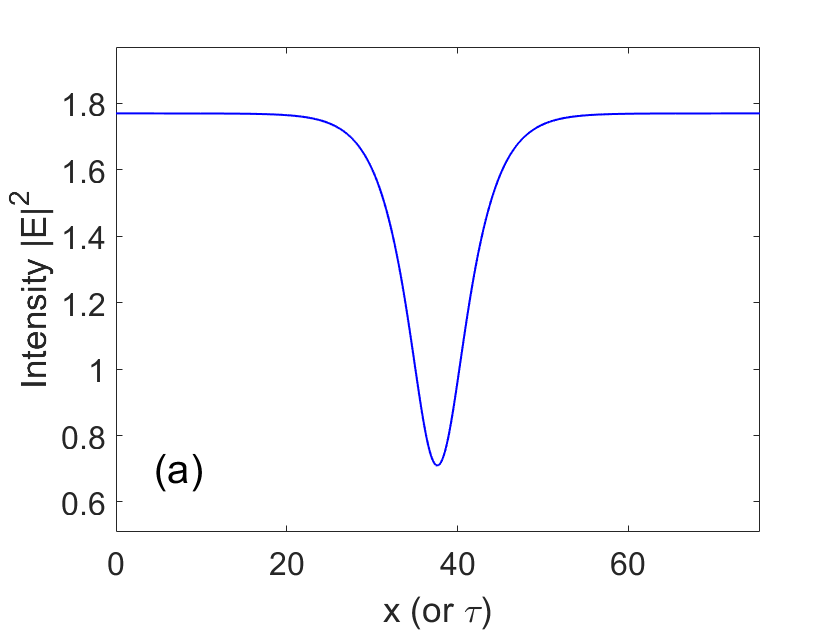}
	\includegraphics[width=0.32\columnwidth]{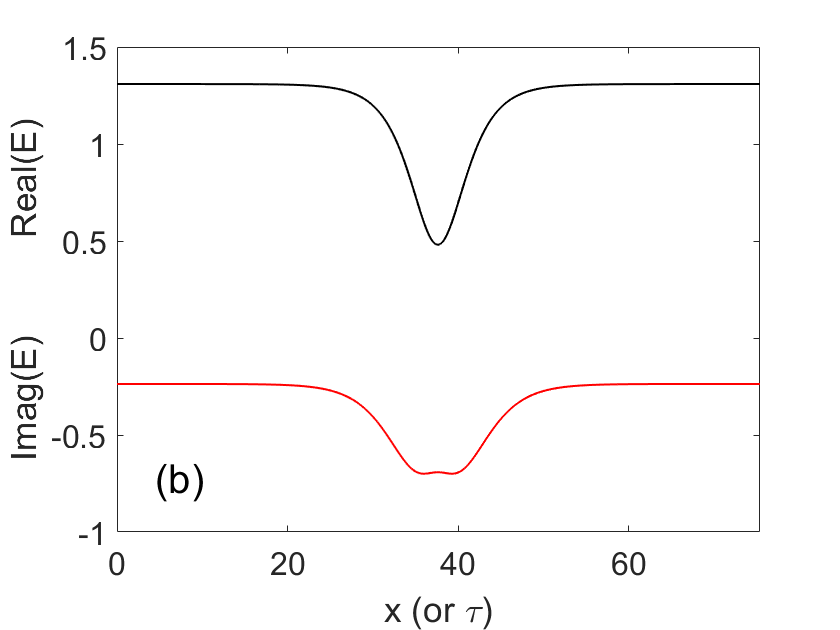}
	\includegraphics[width=0.32\columnwidth]{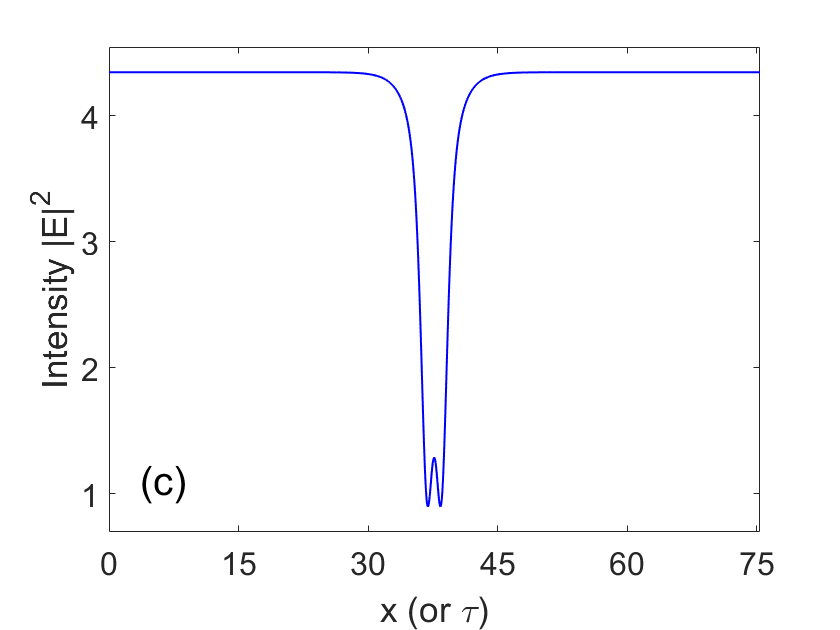}
	\caption{(a) Intensity of a dark CS of the LLE model (\ref{FTLLE}) with normal dispersion for $E_I=1.3515$ and $\theta=1.95$. (b) Real and imaginary parts of the CS field. (c) Intensity of a dark CS of the LLE model (\ref{LLE}) with normal dispersion for $E_I=2.2$ and $\theta=4$.}
	\label{DarkCS}
\end{figure}

Things are different for the dispersive case where, depending on the material used, the group velocity dispersion coefficient can change from positive (anomalous dispersion) to negative (normal dispersion). CSs survive this change but instead of bright CSs, one observes dark (actually grey) CSs in the normal dispersion regime \cite{ParraRivas2016}. Fig.~\ref{DarkCS} shows two examples of dark CSs when the coefficients $a$ or $\beta$ are negative and for parameter values of $E_I=1.3515$, $\theta=1.95$ (panels (a) and (b)), and $E_I=2.2$, $\theta=4$ (panel (c)). When increasing the input amplitude and the detuning, dark CSs develop local oscillations around the trough.  

\begin{figure}
	\includegraphics[width=0.45\columnwidth]{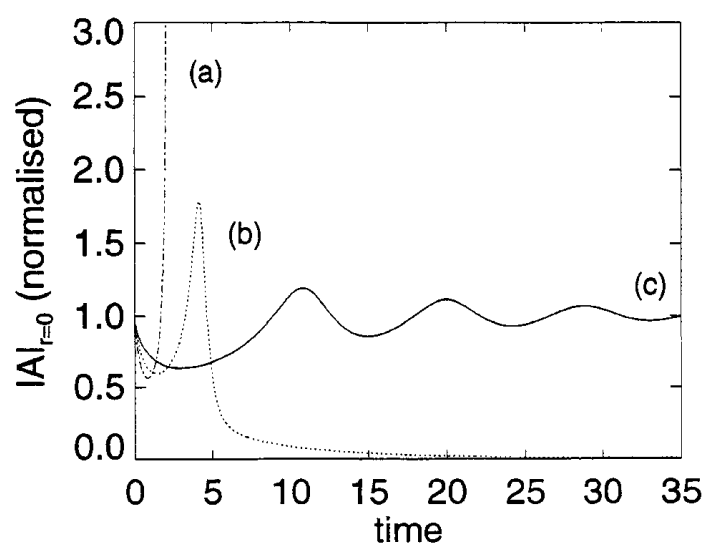}
	\includegraphics[width=0.55\columnwidth]{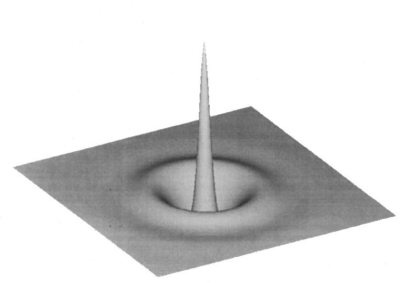}
	\caption{Left panel: Time evolution of a perturbation amplitude in 2D located at the origin for (a) $\theta= 1.8$, collapse to a spatial singularity, (b) $\theta = 1.4$, decay to homogeneous background, and (c) $\theta=1.2$, relaxation to a 2D CS displayed in the right panel. The other parameter is $|E_s|^2=0.9$. Figures reprinted from \cite{Firth1996_PS} and \cite{Firth2002} through authors' permission.}
	\label{CS2D}
\end{figure}

For the diffractive case it is important to mention the two dimensional (2D) case. It is well known that solitons in the NLSE in 2D, a conservative system, undergo a generalised catastrophic collapse \cite{NLSE2005}. This is no longer the case for dissipative systems like the LLE where, however, simulations require due care because of possible effects of boundary conditions. In 1996 three important regimes where identified for the 2D LLE \cite{Firth1996_PS}. The left panel of Fig.~\ref{CS2D} shows the results of numerical simulations of the LLE in 2D for $|E_s|^2=0.9$. For $\theta= 1.8$, for example, we are in the presence of a collapse, much as in the NLSE limit. For $\theta = 1.4$ the peaked solution decays to the flat background. For $\theta=1.2$ damped oscillations of the CS peak indicate asymptotic stability. The profile of an asymptotically stable CS in 2D for $\theta=1.2$ is displayed in the right panel of Fig.~\ref{CS2D} \cite{Firth2002}.

\subsection{Cavity Solitons in Absorptive Media}
\begin{figure}[b]
	\includegraphics[width=0.32\columnwidth]{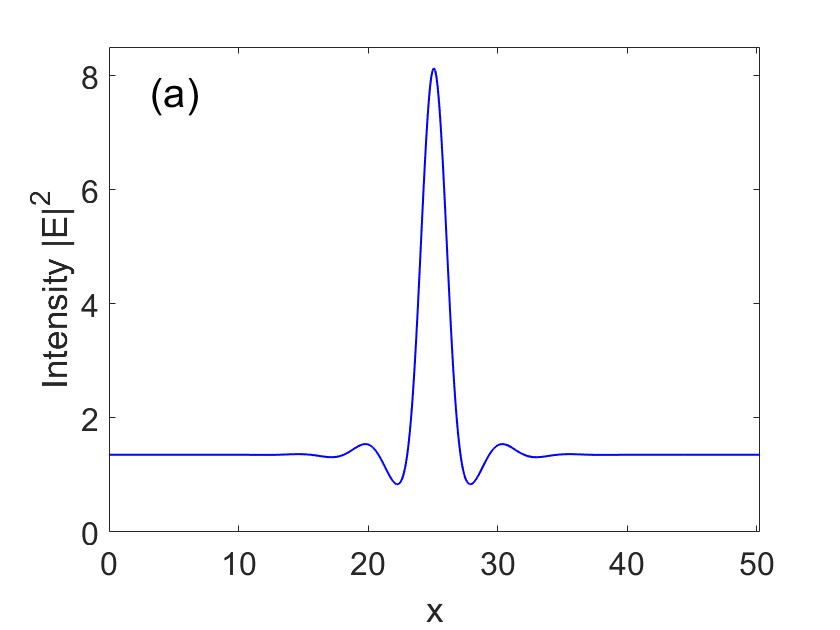}
	\includegraphics[width=0.32\columnwidth]{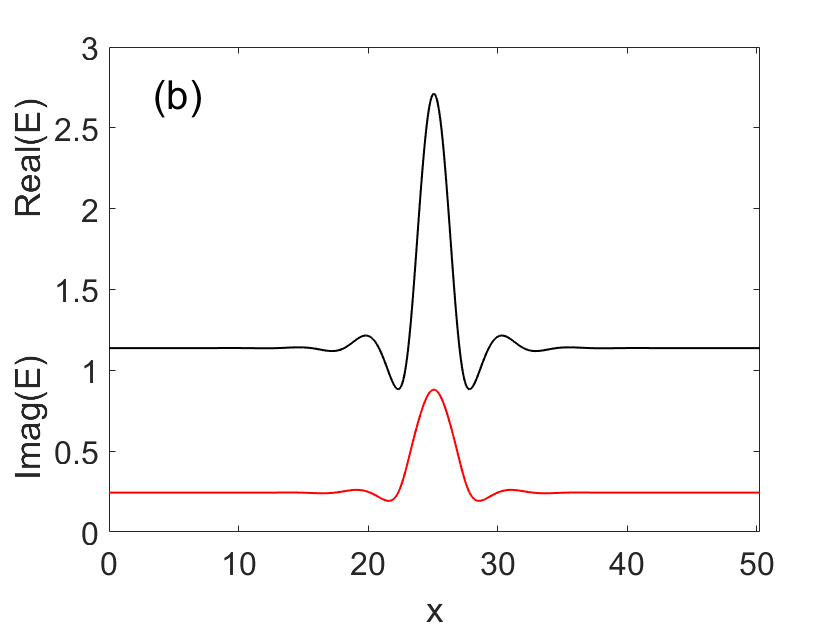}
	\includegraphics[width=0.32\columnwidth]{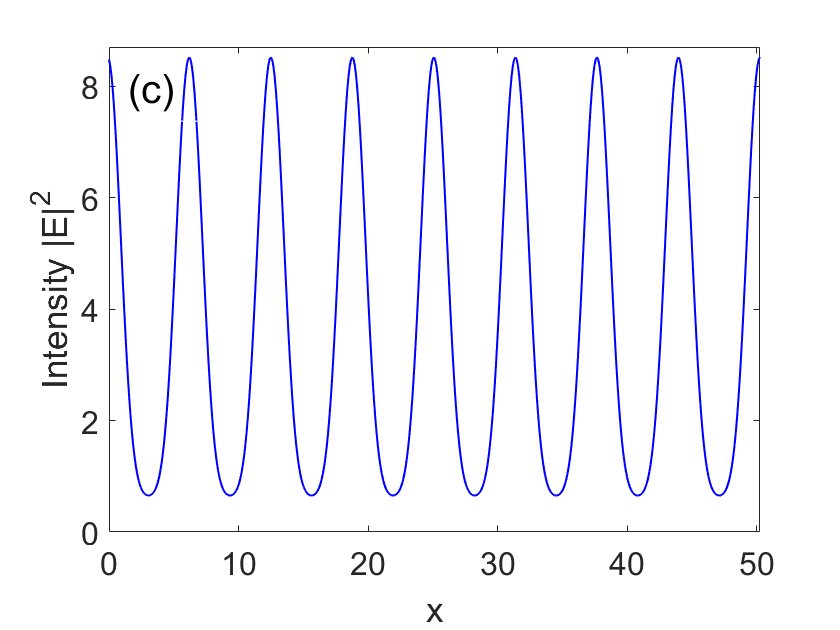}
	\caption{(a) Intensity of a CS of the pure absorptive model (\ref{CavityPureA}) for $\theta=-1.2$, $C=5.4$ and $E_I=6.65$. (b) Real and imaginary parts of the CS field. (c) Stable pattern at the same parameter values.}
	\label{CSPA}
\end{figure}

In the case of a purely absorptive medium, Eq.~(\ref{CavityPureA}), homogeneous stationary states are given by \cite{Firth1996PRL}:
\begin{equation}
E_I^2=|E_s|^2 \left[ \left( 1 + \frac{2C}{1+|E_s|^2} \right)^2 + \theta^2 \right] ,
\label{HSSPA}
\end{equation} 
and, depending on the values of $\theta$ and $C$, the plane-wave input-output characteristic may be either monostable or bistable. We consider CSs in the monostable regime, demonstrating again that they are a phenomenon independent of bistability of homogeneous states. There is a Turing instability for $IS > (S+1)$ where $I=|E_s|^2$ and $S=2C/(1+I)^2$ is a saturation parameter. At threshold, the critical wave vector is $aK_c^2= -\theta$ which is real only if $\theta$ is negative. As for the LLE case, a subcritical condition for Turing patterns is ideal to obtain simultaneously stable patterns and homogeneous stationary states. Following \cite{Firth1996PRL}, we select $\theta=-1.2$, $C=5.4$ and $E_I=6.65$ where the lower branch of the homogeneous solutions is stable and coexists with a stable branch of periodically modulated patterns. Again, we start from a homogeneous input beam $E_I$ with a strong perturbation in its middle. We numerically simulate the Eq.~(\ref{CavityPureA}) with a split-step Fourier method and observe the formation of a CS. After a short transient, the perturbation is removed and the flat input beam restored but the CS solution survives indefinitely as shown in Fig.~\ref{CSPA}. The CS intensity peak is shown Fig.~\ref{CSPA}(a), its real and imaginary parts in Fig.~\ref{CSPA}(b) and a coexisting pattern solution in Fig.~\ref{CSPA}(c). For 2D CSs in the purely absorptive case, see \cite{Firth1996PRL}. Note that CSs in saturable absorbers have been labelled as Optical Bullet Holes (OBH) \cite{Firth1996PRL}. After the perturbations on the input beam have been switched off, these CSs appear as transparent disks on an absorbing background, formed by aiming a short pulse of light at a target location. The existence and dynamical properties of OBH are limited by the decay time of the underlying two-level system. But the strength of the nonlinearity is essentially proportional to that decay time. In the transverse domain, this is manageable, because one can explore the creation, interaction and erasure of OBH on time scales long enough to make the nonlinearity strong. The transverse domain lends itself naturally to semiconductor etalons, where the electronic nonlinearity is much slower than a round-trip time, but correspondingly stronger. Further, one can use Fabry-Perot etalons, which are very compact as well as efficient, with extra nonlinearity due to bidirectionality.
%

\section{Historical Highlights and Properties of Cavity Solitons}
The story of cavity solitons can perhaps be divided into two Ages: the Space Domain Age (from 1994 to 2010), and the Time Domain Age (after 2010). During the Space Domain Age (diffractive models in one or two transverse dimensions) several favourable circumstances combined to stimulate a huge expansion of interest in spatio-temporal nonlinear optics. Workstation development made dynamical simulation of one and two-dimensional field patterns widely available \cite{Harkness2002,McSloy2002}. In addition, simple mean-field models such as Eqs.~(\ref{LLE}), (\ref{CavitySA}) and (\ref{CavityPureA}) provided a convenient framework which readily matched on to studies of pattern formation in fluids and other fields. A special issue of Chaos, Solitons and Fractals in 1994 \cite{Chaos1994} shows how dramatically the field changed in just a few years. It is in this special issue that the first evidence of CSs in the LLE in 1D appears \cite{Scroggie1994} followed by 2D predictions in \cite{Firth1996_PS}, \cite{FirthLord1996} and \cite{Firth2002}. 

For the saturable absorber and purely absorptive models, a breakthrough came in 1996 with the prediction and characterization of optical bullet holes \cite{Firth1996PRL}. Here it was also shown that CSs are very sensitive to spatial variation of the input phase by demonstrating that CSs move up phase gradients and that it is possible to arrange initially randomly distributed CSs into regular grids. If these discrete grids are periodic in space and fully occupied by CSs, one can talk of 'CSs crystals'. Addressing and erasing CSs in the saturating models with input perturbations of suitable phase was also described in the same year in \cite{Brambilla1996}. Control on the spatial position of CSs and their localised switching on and off by using address pulses opened up applications of CSs in optical processing and optical memories. 

On the experimental side two major breakthrough of generation and application of CSs have been achieved at the Institute Non Lineaire in Nice, France (now part of INPHYNI). In 2002 by using a Vertical Cavity Surface Emitting Laser (VCSEL) below threshold when operating as an amplifier, CSs were generated by using a semiconductor medium \cite{Barland2002}. Here it was shown that the ability to switch cavity solitons on and off and to control their location and motion by applying laser pulses makes them interesting as potential ‘pixels’ for reconfigurable arrays or all-optical processing units. A second major application of CSs was demonstrated again in Nice and again by using VCSELs in 2008: an all-optical delay line based on the lateral drift of cavity solitons in semiconductor microresonators \cite{Pedaci2008}. We will see below how these control features are based on intrinsic properties of CSs.

The Time Domain Age (dispersive models in the fast time longitudinal direction of a cavity) developed from 2010, beginning with the seminal work on the experimental realization of CSs in an optical fibre loop \cite{Leo2010,Firth2010}. To separate the diffractive CSs in the transverse plane from the CSs due to group velocity dispersion and propagating along the fibre, the name Temporal Cavity Solitons (TCSs) has been introduced. But the TCSs are still solutions of the LLE and are mathematically identical to the diffractive CSs shown above. The physical properties of TCSs are however unique. In \cite{Leo2010} around 5,000 TCSs were created and sustained in a 300 m long fibre loop (see Fig.~\ref{Cavities}(b)). The key advantage of this set-up is that each TCS propagates around the cavity, sampling and averaging the entire nonlinear medium. This means that the time-translation symmetry of the time-domain LLE is much better observed in fibre loops than the space translation symmetry is in the spatial-domain of media supporting diffractive CSs. 

Around the same time Pascal Del'Haye and his group realised that light propagating in a monolithic ring microresonator was capable to generate an output spectrum with a huge number of discrete lines and a span of over 500 nm ($\approx$ 70 THz) around 1550 nm, a frequency comb, without relying on any external spectral broadening \cite{Pascal2007}. It was later realised first theoretically \cite{Coen13,Chembo13} and then experimentally by Tobias Herr and co-workers \cite{Herr2014} that this new method to generate frequency combs was due to bright TCSs circulating in the Kerr ring resonator. Since then there has been an explosion of theoretical, numerical, experimental and industrial work on frequency combs generated by TCSs of the LLE \cite{Pasquazi2018} of exactly the same kind, shape and stability of those discovered in the early 1990s at Strathclyde \cite{Scroggie1994}. On the theoretical side, it is worthwhile to mention a recent model that unifies TCSs and frequency combs in active and passive cavities \cite{Columbo2021}.

\subsection{Control and Frequency Properties of Cavity Solitons} 
To better understand how this seismic revolution in applications of CSs has taken place, it is important to understand a few properties of CSs: their motion on phase gradients, their controlled positioning (tweezing), their interactions, and their frequency spectra. If we consider an input beam $E_I= E_{I0} \exp[i \phi(x,y)]$ with a space-dependent phase $\phi(x,y)$, the damping and detuning coefficients develop a spatial dependence, and solutions of the LLE (\ref{LLE}) or absorptive equation (\ref{CavityPureA}) acquire a drift velocity given by $\vec{v} = 2 a \vec{\nabla} \phi$ ($v = (2 \beta \partial_\tau \phi$) for the dispersive case) \cite{Firth1996PRL}. As a consequence, a CS will move towards the local maximum of $\phi(x,y)$, and remain there. A pixel array of CSs can be made if $\phi(x,y)$ has an array of maxima. This was demonstrated in \cite{Firth1996PRL} by writing the letters 'IT' with CSs in a square array of phase maxima of the input beam. The process of moving and controlling final positions of CSs is similar to optical tweezers where laser light is used to trap and move microscopic particles in space. The difference is that here light is capable of tweezing CSs. In 2015 many TCSs were stored in an optical fibre loop pumped by a continuous wave ‘holding’ laser beam \cite{Jang2015}. The cavity solitons are trapped into specific time slots through a phase modulation of the holding beam, and moved around in time by manipulating the phase profile. Continuous and discrete tweezing and manipulations of the temporal positions of TCSs were achieved experimentally, with the ability to simultaneously and independently control several pulses within a train.

\begin{figure}
	\includegraphics[width=0.68\columnwidth]{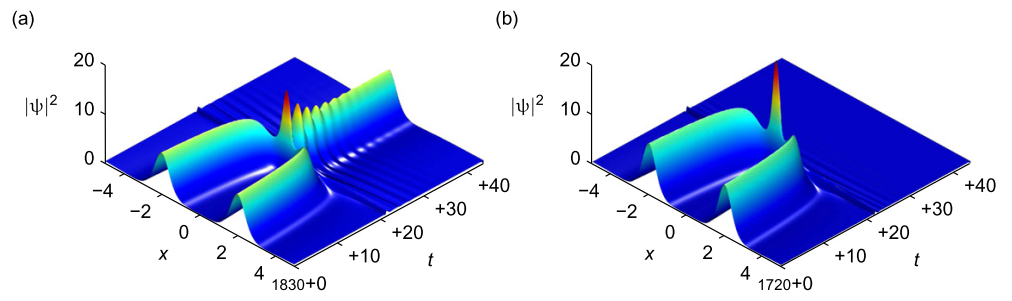}
	\includegraphics[width=0.30\columnwidth]{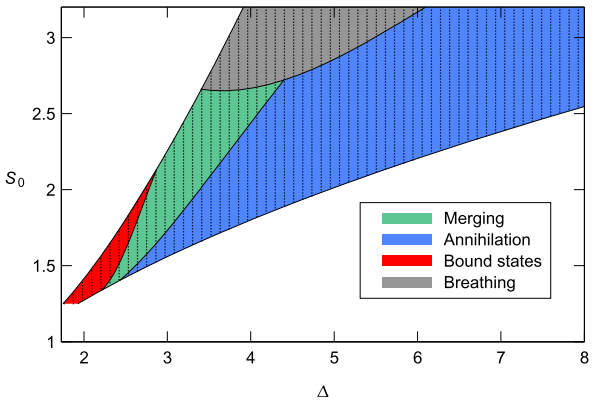}
	\caption{Numerically simulated dynamics of induced TCSs interactions for $\phi(0) = 0.5 rad$ and $x_0 = 30$ using the LLE (\ref{LLE_16}). In (a) for $S=1.87$, $\Delta=2.91$, the two TCSs merge into one; in (b) for $S=2.10$, $\Delta=3.64$, the two TCSs annihilate one another. Right panel: Results from numerical simulations illustrating the outcome of TCSs interactions as a function of the detuning $\Delta$ and driver strength $S_0$. Each solid dot represents a distinct simulation. No TCS exists in the white area. Figures reprinted from \cite{Jang2016} through authors' permission.}
	\label{Merge1}
\end{figure}

Conventional solitons of conservative integrable systems like those in Kerr media described by the NLSE (\ref{NLSE}) pass through each other without radiation losses \cite{Zakharov1972}. This follows from the fact that the NLSE has an infinite number of integral invariants. Things change of course when considering non-Kerr materials as for example in \cite{Snyder1993} where non--NLSE solitons radiate on colliding with the radiation creating new stable solitons or fuse the original two into a single stable soliton. In contrast, theoretical and experimental observations of soliton collisions in the LLE (\ref{LLE}), a NLSE with the addition of input and loss of energy, have been far rarer. Unlike conservative solitons, widely separated CSs of LLE are phase-locked to the external driver, and thus all of them possess identical features (for given $E_I$ and $\theta$), including frequency and velocity. Accordingly, unassisted collisions occur only when two CSs are sufficiently close to interact with each other \cite{Brambilla1996}, yet such interactions are difficult to explore controllably. Inducing collisions by suitably modulating the phase of the driver addresses that issue \cite{McIntyre2010}. Since a CS moves towards the local maximum of a background phase modulation, a controlled collision of CSs can be observed when exciting, for example, two CSs on opposite sides of a local maximum of the phase profile \cite{McIntyre2010}. Controlled collisions of TCSs leading to merging and annihilation have been predicted and observed in a detailed numerical and experimental study of the LLE in \cite{Jang2016}. In \cite{Jang2016} the following form of the LLE for the field $\psi$ has been used:
\begin{eqnarray}
\partial_{t} \psi = S -(1+i\Delta) \psi + i  |\psi|^2 \psi + i \partial^2_\tau \psi . 
\label{LLE_16}
\end{eqnarray}
where $S$ is the amplitude of the input field and $\delta$ the cavity detuning.
In both merging and annihilation cases the dissipative nature of the interaction is revealed through the analysis of the energy balance. Fig.~\ref{Merge1} shows the results of numerical simulations of the LLE (\ref{LLE}) with an input beam $E_I=E_{I0} \exp(i \phi(x))$ with a Gaussian phase profile $\phi(x)=\phi(0) \exp (-x^2/x_0^2)$. Here the longitudinal spatial coordinate $x$ is related to the fast time variable $\tau$ of Eq. (\ref{FTLLE}). Two TCSs are initiated through the initial condition
\begin{equation}
E(0,x) = \sqrt{2 \theta} \left[ {\rm sech}\left(\sqrt{\theta} (x-x_L) \right) + {\rm sech}\left(\sqrt{\theta} (x+x_L) \right) \right]
\end{equation}
where $2 x_L$ is the initial distance of the TCSs. Figs.~\ref{Merge1}(a) and (b) show typical results of the LLE simulations for $E_{I0}=1.87$, $\theta=2.91$ and $E_{I0}=2.10$, $\theta=3.64$ and with $\phi(0) = 0.5 rad$ and $x_0 = 30$. These parameters are chosen to replicate the experiments described below. In both cases, the TCSs drift towards each other until they are close enough to interact. The outcome of the collision is, however, markedly different. In Fig.~\ref{Merge1}(a) the two TCSs merge into one, while in Fig.~\ref{Merge1}(b) the intracavity field after the interaction is globally reduced to the homogeneous solution, i.e. the two TCSs annihilate one another. In the left panel of Fig.~\ref{Merge1}, a summary of the numerically observed outcome of the interactions as a function of $E_{I0}$, $\theta$ for the same driver phase modulation $\phi(0)$ and $x_0$ is presented. Merging (green) and annihilation (red) of TCSs occur in clearly distinct, but adjacent, regions \cite{Jang2016}.
\begin{figure}
	\includegraphics[width=0.30\columnwidth]{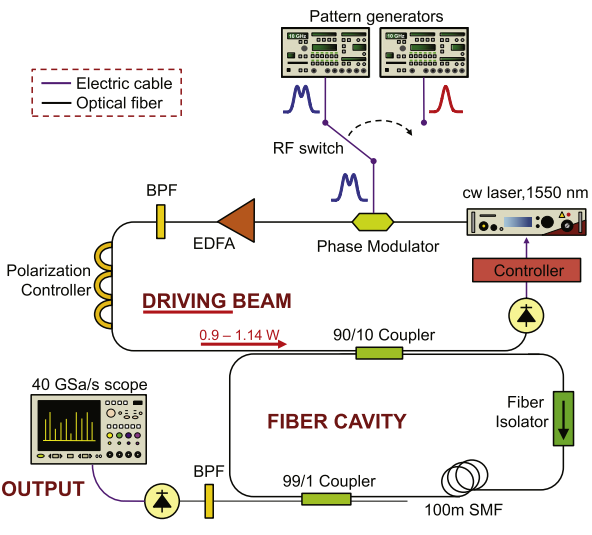}
	\includegraphics[width=0.68\columnwidth]{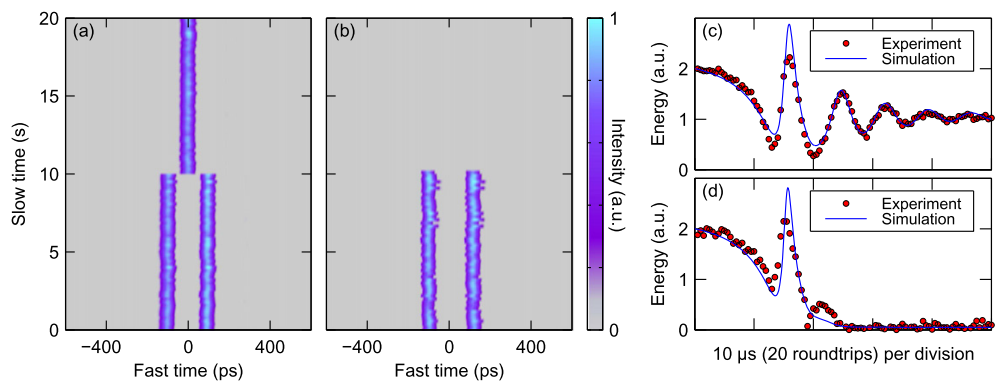}
	\caption{Left panel: Experimental setup. cw: continuous-wave, EDFA: erbium-doped fibre amplifier, BPF: band-pass-filter, SMF: single-mode fibre. (a),(b) Experimental plots showing the evolution of the intra-cavity temporal intensity profile as two TCSs (a) merge into one, and (b) annihilate each other. (c),(d) The roundtrip-to-roundtrip evolution of the total intra-cavity energy during TCSs (c) merging and (d) annihilation. The experimental data is represented by red circles, and results from numerical simulations of the LLE are shown as blue solid lines. Figures reprinted from \cite{Jang2016} through authors' permission.}
	\label{Merge2}
\end{figure}

Experiments were performed in a nonlinear optical fibre (Kerr) resonator (see the left panel of Fig.~\ref{Merge2}), in which the TCSs are excited at selected and precise positions, and systematically induced to interact. As for the numerical simulations, the interactions are triggered by manipulating the phase profile of the driver; the outcome controllably depends on the detuning $\theta$ and driving strength \cite{Jang2016}. Typical results are shown in Fig.~\ref{Merge2}(a) and Fig.~\ref{Merge2}(b) for parameter conditions close, to experimental accuracy, to those of Fig.~\ref{Merge1}(a) and Fig.~\ref{Merge1}(b), respectively. The first 10 s of the measurements are very similar: two TCSs with 200 ps separation are stably trapped at the maxima of the phase pulses. In Fig.~\ref{Merge2}(a) the experimental merging to a single TCS is observed while both TCSs disappear in an annihilation event in Fig.~\ref{Merge2}(b). These results are strongly indicative of merging and annihilation, which is in agreement with numerical simulations. Indeed, the simulation results use the very same parameters as the experiments \cite{Jang2016}.

To clearly establish the origin of the observed dynamics, the roundtrip-by-roundtrip evolution of the intracavity energy was measured on a real-time oscilloscope. Typical experimental results for merging and annihilation are shown as red circles in figures Fig.~\ref{Merge2}(c) and (d), respectively \cite{Jang2016}. The energy is normalised so that a single isolated TCS carries an energy of 1 in arbitrary units. The results unambiguously reveal the dissipative nature of the interactions: for the case of Fig.~\ref{Merge2}(a) the energy falls from two to one, implying merging; for Fig.~\ref{Merge2}(b) the energy falls from two to zero, implying annihilation. The agreement between experiments and numerical simulations is simply spectacular, a testament to the accuracy and predictive power of the equation models considered here. The localized and dissipative nature of CSs is clearly demonstrated theoretically, numerically and experimentally and confirms in an unequivocal way analogies and differences of CSs from conservative Kerr solitons of the NLSE.

\begin{figure}
	\includegraphics[width=0.24\columnwidth]{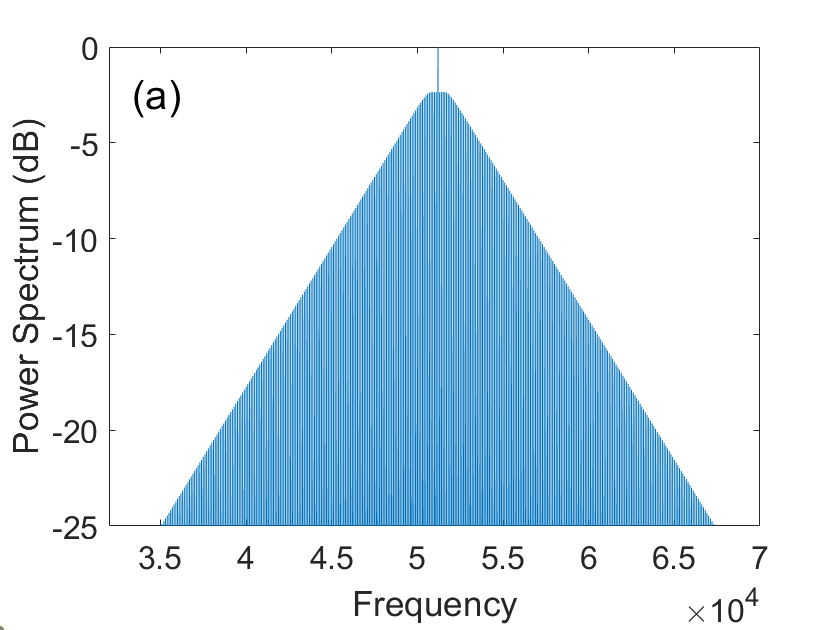}
	\includegraphics[width=0.245\columnwidth]{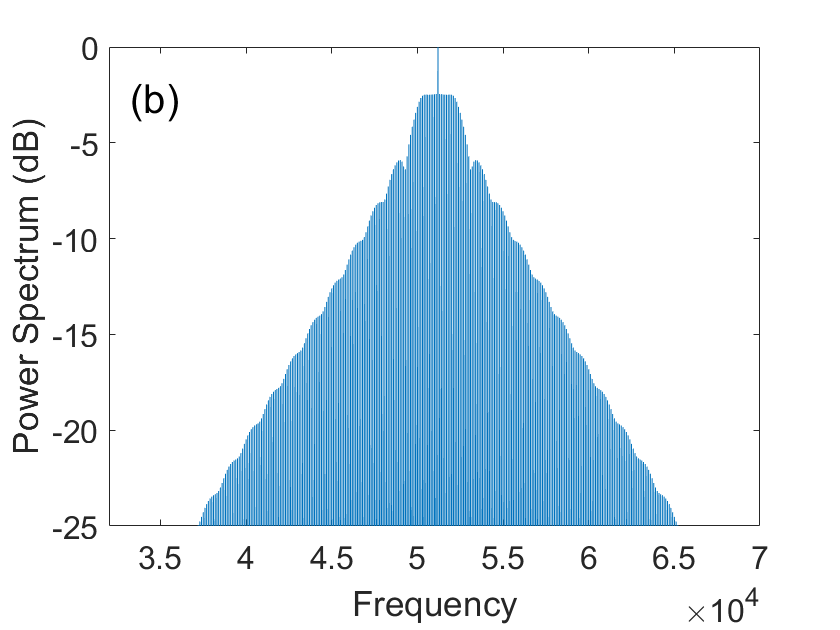}
	\includegraphics[width=0.245\columnwidth]{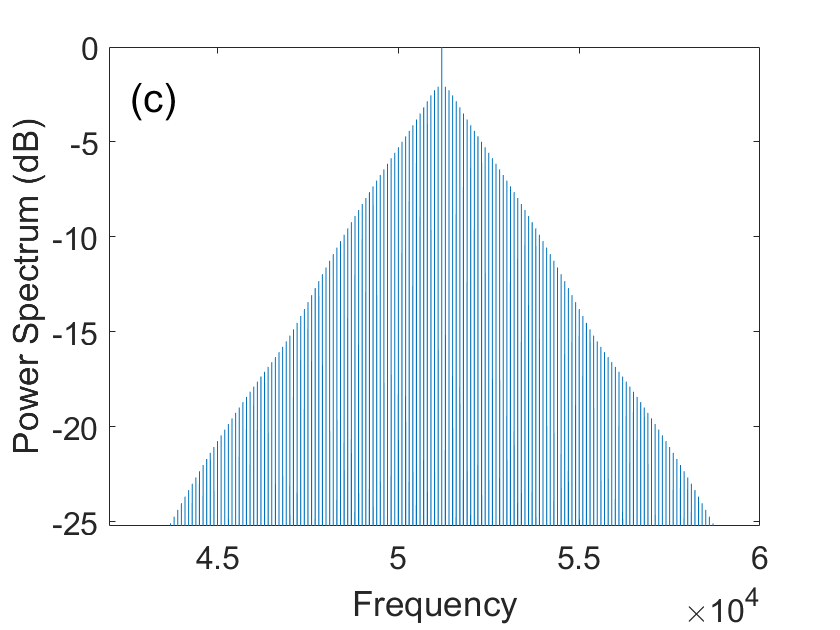}
	\includegraphics[width=0.245\columnwidth]{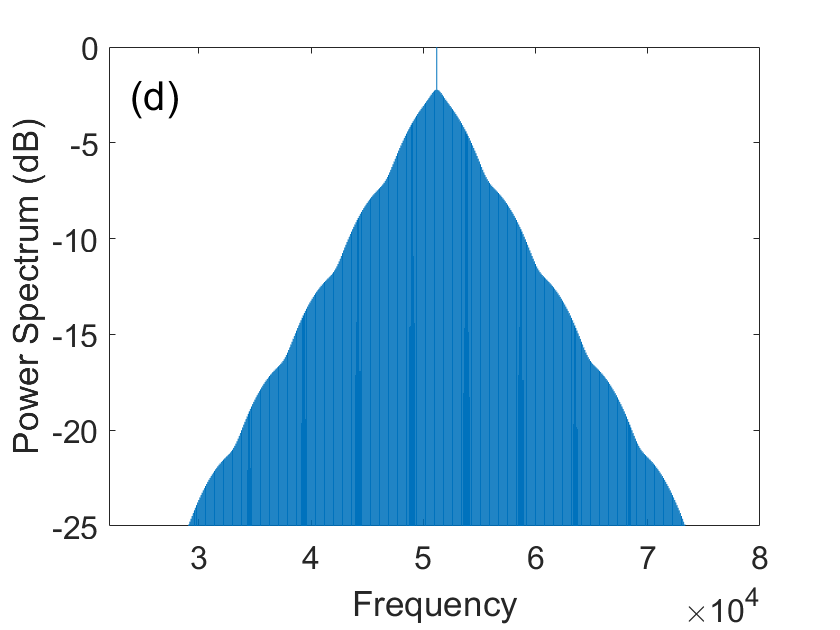}
	\caption{Frequency power spectra of CSs for (a) the bright LLE CS of Fig.~\ref{CS}(a), (b) the bright purely absorptive CS of Fig.~\ref{CSPA}(a), (c) the dark LLE CS of Fig.~\ref{DarkCS}(a) and (d) the dark LLE CS of Fig.~\ref{DarkCS}(c).}
	\label{Spectra}
\end{figure}

Finally, we consider the spectral properties of TCSs. The output of photonic devices generating TCSs is shown schematically in Fig.~\ref{Cavities}(b). A c.w. input laser produces an output of regularly spaced optical pulses, the CSs. In Fig.~\ref{Spectra} we show the power spectrum in decibels of a train of bright LLE CSs of Fig.~\ref{CS}(a)-(b), of a train of bright purely absorptive CSs of Fig.~\ref{CSPA}(a) and Fig.~\ref{CSPA}(b), and of a train of the two dark LLE CSs of Fig.~\ref{DarkCS}(a)-(c), respectively. 
There are three clear common features in these four power spectra: 1) Broadness that survives to very low decibel scale; 2) Finely spaced comb teeth separated by the inverse of the round trip time of the micro-resonator; 3) Almost pyramidal structure with a single central peak. These are the features that have made CSs the optimal generators of optical frequency combs. Frequency combs consist of equidistant laser lines, and have revolutionized time-keeping, metrology, and spectroscopy. Hence, the investigation of power spectra from devices producing TCSs has become an essential tool for the applications reviewed in the next section. Here we note that the original LLE CS of \cite{Scroggie1994} produces a very broad spectrum with almost exact triangular shape and a huge number of comb lines, see Fig.~\ref{Spectra}(a). The spectrum of the purely absorptive case, Fig.~\ref{Spectra}(b), has similar features to the LLE case but the sides are modulated due to a fast-time varying phase that can be seen in Fig.~\ref{CSPA}(b). Finally the intensity modulations of the dark LLE CS of Fig.~\ref{DarkCS}(c) are clearly visible on the side of the spectrum in Fig.~\ref{Spectra}(d) when compared to the spectrum in Fig.~\ref{Spectra}(c) corresponding to the unmodulated dark LLE CS of Fig.~\ref{DarkCS}(a).

\section{Applications of Cavity Solitons}
In this section we review some of the most striking applications of CSs in photonic devices. In their thirty years of life the number of experimental observations and applications has grown enormously with a recent explosion in scientific impact after the realization of TCSs in 2010 \cite{Leo2010}. 

\subsection{Optical Memory Based on Cavity Solitons}
Beyond their fundamental interest, CSs have been sought in optical resonators as elementary bits for information processing. The idea is to use the transverse area of an optical resonator as a blackboard, where light bits can be individually written and erased, thus forming reconfigurable optical memory arrays \cite{McDonald1990,Firth1996PRL,Lugiato2003}. In these devices, address pulses of short duration can imprint CSs in specific locations in the plane perpendicular to the direction of propagation in the cavity \cite{Barland2002}. We have seen that these CSs can be moved to desired positions, say on a square matrix, by suitable modulations of the input phase \cite{Firth1996PRL}. In this way messages can be coded and stored inside the cavity for as long as one desires, thus building an optical memory. The messages can be changed at will by using new address pulses with suitable phase \cite{Brambilla1996}. Hence the realization of reconfigurable optical memories \cite{Pedaci2006}. Because of the limited size of photonic devices in the transverse plane and the the typical size of CSs being around 10$\mu$m, optical memories based on diffractive CSs to date are of the order of a hundred bits of data (8x8 matrices of CSs). Things are different when considering TCSs in long fibre loops (380 m) driven by a c.w. laser at 1550 nm \cite{Leo2010}. TCSs of 4 ps duration were generated via ‘addressing’ pulses. The TCSs circulated unchanged for hundreds of thousands of cavity round trips, through balances of dispersion and nonlinearity, on one side, and driving and losses, on the other. An all-optical memory and buffering of 15-bit data streams were achieved \cite{Leo2010,Firth2010}. In 2015, by applying suitable phase modulations to avoid TCS interactions, a robust all-optical buffer based on TCSs and capable of storing a record 4536 bits of data at 10 Gb/s was demonstrated \cite{Jang2016OM}.

\subsection{Optical Information Processing Based on Cavity Solitons}
Maximising information capacity in optoelectronics demands some kind of buffering or delaying mechanism for avoiding the so-called data packet contention in data transmission. This problem originates from switches or routers that can only process one packet at a time. The solution is to build a buffer, which places one of the packets on hold while the other clears the switch. Several methods have been suggested to realize an all-optical delay line, the most famous being slow light \cite{Boyd2005}. Slow-light schemes however are usually limited to a few pulse widths. 

As mentioned in Section 4, an alternative approach to an all-optical delay line was developed using semiconductor CSs \cite{Pedaci2008,McIntyre2010}. This method injects a stream of optical bits into an optical cavity, creating CSs that drift transversely with a controllable velocity. A phase gradient imposed on the input input holding beam, induces drift which removes the CSs from their initial positions and moves them toward the maxima of the gradient. The speed at which the CSs move and the resulting delay, depend on the form and strength of the applied gradient \cite{Pedaci2008,McIntyre2010}. Recent extensions of this method to semiconductor lasers (VCSELs) with saturable absorbers \cite{Eslami2014} suggested less-complex configurations, possible realizations in integrated circuits and the remarkable capability of tuning the delay up to at least 2300 pulse widths. These features make delay and buffering methods based on CSs viable technologies in optical processing of information and all-optical computing. 

\subsection{Frequency Combs Based on Cavity Solitons} 
Conventional optical frequency combs based on mode-locked laser pulses in photonic crystal fibres \cite{HanschHall2006} are still mostly confined to scientific laboratories \cite{Fortier2019}. In recent years, there has been progress in the development of optical frequency combs based on compact, chip-scale microresonators (also named micro-combs), based on CSs \cite{Herr2014}. Micro-combs are now capable of producing coherent, octave-spanning frequency combs, with microwave to terahertz repetition rates, at low pump power, and in chip-scale devices and have been used in a wide variety of applications, owing to bandwidth and coherence provided by the dissipative temporal soliton states. Here we list here some of the most striking applications of micro-combs based on CSs, the vast majority of them already in operation. 

{\bf Optical communications.} 
Replacing a large number of lasers in wavelength-division-multiplexed (WDM) optical communication systems with an optical frequency comb has been an attractive idea for some time \cite{Takara2000}. Optical frequency combs have an intrinsically stable frequency spacing that enables transmission-performance enhancements beyond what is possible with free-running lasers. Moreover, by using a frequency comb in WDM systems one can relax the resource requirements at the receiver by implementing joint impairment compensation and tracking for multiple data channels by exploiting the broadband phase coherence of the frequency comb. Micro-combs (or Kerr combs) use the Kerr effect in an integrated microcavity to convert light from a continuous-wave pump laser to evenly spaced lines across a wide bandwidth \cite{Pascal2007,Herr2014}. The performance of micro-combs is sufficiently high to cope with the requirements in terms of frequency stability, signal-to noise ratio, and linewidth of modern coherent communication systems \cite{Pfeifle2014,Liao2017,Fulop2017}. Stabilized Kerr CSs in microresonators guarantee control of the bandwidth and number of comb lines with great precision \cite{Marin2017}. By using two SiN micro-combs, thermal control and tuning of the central frequency has allowed the use of a matched comb at the receiver as a multi-wavelength local oscillator \cite{Marin2017}. The line spacing can reach values in the order of 100 GHz, making micro-combs an attractive multi-wavelength light source for applications in fiber-optic communications. Micro-combs based on dark CSs \cite{Xue2015} are particularly interesting for advanced coherent communications since they can display lower time jitter, e.g. 13 dB lower \cite{Lao2023}, than bright CSs. For example, a coherent-transmission line using 64-quadrature amplitude modulation encoded onto the frequency lines of a dark CS comb has enabled transmitted optical signal-to-noise ratios above 33 dB in an 80-km data transmission with 20 channels \cite{Fulop2018}.

{\bf Optical clocks, frequency standards, GPS.}
Optical frequency combs have revolutionized metrology and advanced other fields such as RF photonics and astronomy. As mentioned above, traditional frequency combs can be bulky, expensive, and difficult to manufacture, thus limiting their use in real-world scenarios. Within the last decade or so, CS based micro-combs have led to hopes of overcoming the constraints of more traditional bulk combs.

One of the first applications for bulk frequency combs was the optical atomic clock. It promised extreme long-term time stability, better than that of the Cesium clock that currently defines the SI second. More recently, interest in a fully portable optical atomic clock has grown. Such a device could reliably keep time even without the aid of GPS references, and potentially with greater accuracy than current GPS synchronization can provide. Optical clocks take advantage of narrow and stable atomic transitions to realize exceptionally stable laser frequencies \cite{Hinkley2013}. Optical frequency combs facilitate the measurement and use of these atomic references by providing a set of clock-referenced lines that span more than an octave. Micro-combs offer revolutionary advantages over existing comb technology, including chip-based photonic integration, large comb-mode spacings, and monolithic construction with small size and low power consumption.
Micro-comb technology has advanced frequency control of spectra via the implementation of phase-locked and mode-locked states and even through a Rb-stabilized micro-comb oscillator \cite{Savchenkov2013}. The milestone of all-optical frequency control of a micro-comb locked to to an atomic reference, including frequency division to the microwave domain, has been achieved in \cite{Papp2014} by demonstrating a functional optical clock based on full stabilization of a micro-comb to atomic Rb transitions. 

Even more recently, CSs in Kerr microresonators have been used as a source of coherent, ultrafast pulse trains and ultra-broadband optical-frequency combs to enable optical synthesis and metrology \cite{Drake2019}. Here a Kerr microresonator optical clockwork, distributing optical-clock signals to the mode-difference frequency of a comb has been developed. The clockwork is based on a silicon-nitride microresonator that generates a CS frequency comb with a repetition frequency of 1 THz. These experiments have reached a record absolute frequency noise of one part in $10^{17}$, the highest accuracy and precision ever reported in optical clockworks with the possibility of measuring high-performance optical clocks with Kerr micro-combs.

{\bf Astrocombs.}
Astrocombs are broadband, high-repetition rate optical frequency combs that are used for the calibration of astronomical spectrographs \cite{Steinmetz2008}. Their precision and accuracy make astrocombs a critical technology for astronomical spectroscopy and will likely enable ground-breaking observations in the fields of exoplanets, cosmology and fundamental physics. Conflicting requirements of comb line spacing (usually more than 10 GHz), broadband spectral coverage (from below 400 to above 2400 nm) and low-maintenance operation have been significant technical challenges. For in-depth discussions of astrocombs, see the comprehensive reviews \cite{Herr2019,Roztocki2019,McCracken2017}.

For astrocombs, TCSs in Kerr resonators enable the generation of low-noise, ultra-short femtosecond pulses with repetition rates that can readily reach and exceed tens of GHz. Microresonator combs have been employed in two proof-of-concept demonstrations of near-infrared astronomical spectrograph calibration. A TCS micro-comb generated in a fused silica chip-based resonator whose line spacing of 22 GHz was stabilized and whose free-drifting offset was tracked, was used for calibration of the NIRSPEC spectrograph \cite{Suh2019}. Simultaneously, in a parallel demonstration a 24 GHz TCS comb, fully stabilized via pulsed driving \cite{Obrzud2017}, was generated in a silicon nitride photonic-chip microresonator and used for calibration of the GIANO-B spectrograph\cite{Obrzud2019}. Recently, pulsed driving has enabled demonstrations of octave-spanning near-infrared spectra hence showing that coverage of the entire near-infrared is possible \cite{Li2017,Pfeiffer2017}. Finally, dark TCS combs operating in the normal dispersion regime \cite{Xue2015} provide new opportunities in view of their potential integration in space-based observatories where compactness and low power consumption are key ingredients.

{\bf Quantum technologies.}
We come now to the research topic that would have interested Rodney Loudon the most: applications of CS frequency combs in quantum optics. When a Kerr resonator is pumped weakly, frequency combs can be a quantum resource for the generation of heralded single photons and energy-time entangled pairs \cite{Grassani2015}, multiphoton entangled states \cite{Reimer2016,Kues2017} and squeezed vacuum \cite{Vaidya2020,Zhao2020}. When pumped more strongly, the modes of a Kerr comb can become phase-locked to form a stable, low-noise dissipative Kerr soliton, the TCS. Although the TCS micro-comb has been extensively studied classically (see above), it is nonetheless fundamentally governed by the dynamics of quantized processes: each resonator mode is coupled to every other mode through a four-photon interaction. When this quantum processes can be harnessed, TCS micro-combs can open a pathway toward the experimental realization of a multimode quantum resource in a scalable, chip-integrated platform. Fully quantum-optical properties of TCS micro-combs have now been directly observed first in the quantum-limited timing jitter and quantum diffusion of the TCS state \cite{Bao2021} and then in quantum correlations of a multimode Gaussian state leading to the prediction of an all-to-all entanglement for this state \cite{Guidry2022}.

\section{Conclusions}
Cavity Solitons, a class of dissipative localized solutions observed in photonic devices when coherent light propagates through a Kerr nonlinear medium in an optical cavity, have gone a long way from their first prediction thirty years ago at the University of Strathclyde \cite{Scroggie1994}. By using the Lugiato-Lefever model, it was discovered that CSs balance diffraction (or dispersion) with the Kerr nonlinearity, as happens for conservative solitons in the NLSE, but also the cavity losses with the energy input of the laser driver, a feature that is typical of CSs \cite{Firth1998}. Typical CSs in Kerr or purely absorptive media require Turing instabilities leading to pattern solutions that are bistable with homogeneous stationary states. The tails of CSs approach asymptotically these homogeneous states while the peaks are very close to a single peak of the coexisting pattern solution (see Figs.~\ref{CS} and \ref{CSPA}). All these CSs features, first investigated in the diffractive case of the LLE, survive unaltered in the Kerr micro-resonator \cite{Herr2014} and fibre loops \cite{Leo2010} cases with group velocity dispersion instead of diffraction, the TCSs. This is the important message of this review paper. There is no need to introduce new names like 'dissipative Kerr solitons' \cite{Kippenberg2018,Lugiato2018} to describe well established CSs of the LLE. 

CSs and TCSs are universal features in nonlinear, quantum optical and photonic systems and have been found in VCSELs \cite{Barland2002}, in lasers with saturable absorbers \cite{Bache2005,Elsass2010}, in the presence of higher order dispersion \cite{Tlidi2010,Tlidi2013,Parra2017}, of vectorial cases with two-polarizations \cite{Xu2021}, of quadratic nonlinearities \cite{Oppo2001,Englebert2021}, of three level media with two drivings and electromagnetically induced transparency \cite{Oppo2022} and of laser systems \cite{Tanguy2008,Bao2019} to cite few examples. Dissipative solitons are also found in Fabry-Perot cavities \cite{Cole2018,Campbell2023} and single mirror feedback optical configurations \cite{Ackemann2005,Ackemann2009}.

Since the experimental demonstration of TCSs \cite{Leo2010}, the number of applications of photonic devices using CS based optical frequency combs has exploded: wave demultiplexing in optical communications, frequency standards, optical clocks, future GPS, astrocombs and quantum optic technologies based on micro-combs are just few of the examples that we have described in Section 5. We expect these to continue to grow in the coming decades as technologies for integrated devices further develop. CS based micro-combs are ideal for the processes of miniaturization and device integration.  

\section{Acknowledgements}
During the years we have benefited from research collaborations with Andrew Scroggie, Graeme Harkness, Alison Yao, Thorsten Ackemann, Graham McDonald, Giampaolo D'Alessandro, Damia Gomila, Dmitry Skryabin, Angus Lord, Graeme Campbell, Lewis Hill, Craig McIntyre, Stephane Coen, Miro Erkintalo, Stuart Murdoch, Julien Fatome, Luigi Lugiato, Giovanna Tissoni, Franco Prati, Massimo Brambilla, Alessia Pasquazi, Pascal Del'Haye and many others. We would also like to thank the European Union for the financial support provided via several research networks (PIANOS, FunFACS, ColOpt, QStruct, QuantIM) and grant C299792458MS while also thanking our co-workers on these projects, too many to be named here. 

\section{Appendix}
Starting from Eqs. (\ref{NLSE}) and (\ref{BC}), we complete here the derivation of the LLE in the standard Mean Field Limit (MFL) approximation. In order to force the boundary condition (\ref{BC}) into the propagation equation (\ref{NLSE}), the usual MFL transformation is entered
\begin{eqnarray}
\label{MFLT}
z' &=& z \nonumber \\
t' &=& t + \left [ \frac {\Lambda - L} {c} \right] \frac {z} {L} \,\,.
\end{eqnarray}
Under the condition (\ref{MFLT}), we obtain
\begin{eqnarray}
\partial_z + \frac {n}{c} \,\, \partial_t =  \partial_{z'} + 
\left [ \frac {\Lambda - L} {cL} \right] \partial_{t'} + 
\frac {n}{c} \,\, \partial_{t'} = \partial_{z'} + 
\left [ \frac {\Lambda + (n - 1) L}{cL} \right ] \,\, 
\partial_{t'} \, . 
\end{eqnarray}
By introducing the new field variable $F$ such that
\begin{eqnarray}
F = \Gamma E + \sqrt{T} E_{I} \frac {z}{L} \,\,\,\, 
{\rm with} \,\,\,\, 
\Gamma = \exp \left(D \frac {z}{L} \right )
\end{eqnarray}
we obtain
\begin{eqnarray}
& & \partial_{t'} F + \frac {cL} {\Lambda + (n - 1) L} \;\;\; 
\partial_{z'} F \nonumber = \\
~ & & \frac {cL} {\Lambda + (n - 1) L} \left [ \frac{D}{L} 
\left(F - \sqrt{T} E_{I} \frac {z}{L} \right) + \Gamma 
\left( i \frac{1}{2k} \nabla^2 E + i \eta |E|^2 E \right) + \sqrt{T} E_{I} \frac {1}{L} \right ] .
\end{eqnarray}
The longitudinal boundary conditions (\ref{BC}) are now transformed into 
\begin{eqnarray}
F(x,y,0,t') = F(x,y,L,t') \label{NBC}
\end{eqnarray}
providing a condition of periodicity for the field at the same time $t'$ (synchronous boundary conditions). Under the MFL assumptions of $\Theta$ and $1/(2k)$ being $O(\varepsilon) \ll 1$ one obtains:
\begin{eqnarray}
D \approx - \frac{T}{2} - i \Theta + 
i \frac {\Lambda - L}{2k} \nabla^2 \;\;\;\;\;\;\;\;\;\;\;\;
\Gamma \approx 1 + \frac {D} {L} z 
\end{eqnarray}
since
\begin{eqnarray}
{\rm ln} \sqrt{R} = {\rm ln} \sqrt{1-T} \approx {\rm ln} 
\left( 1- \frac{T}{2} \right) \approx - \frac{T}{2} \; .
\end{eqnarray}
At the first order in $\varepsilon$ one gets:
\begin{eqnarray}
& & \partial_{t'} F + \frac {cL} {\Lambda + (n - 1) L} \,\,\, 
\partial_{z'} F = \nonumber \\ 
& & -\frac {c \; T /2 } {\Lambda + (n - 1) L} \,\, F 
- i \frac {c \; \Theta} {\Lambda + (n - 1) L} \,\, F_j 
+ i \frac {c (\Lambda-L)} {2 k [\Lambda + (n - 1) L]} 
\nabla^2 F \nonumber \\
& & +\frac {c \; \sqrt{T}  } {\Lambda + (n - 1) L} 
\,\, E_{I} + \frac {c \; L} {\Lambda + (n - 1) L} 
\,\, i \eta |E|^2 E .
\end{eqnarray}
By introducing the new convenient parameters
\begin{eqnarray}
\label{NPAR}
\tau = \frac {\Lambda + (n - 1) L}{c} ; \;\;\;\;\;\;\;\;
d = \frac {\Lambda-L} {2 k};
\end{eqnarray}
one obtains:
\begin{eqnarray}
\tau \,\, \partial_{t'} F + L \,\, \partial_{z'} F = 
- (T/2) F - i \Theta F + i d \nabla^2 F + i L \eta |E|^2 + \sqrt{T} E_{I} \, . 
\end{eqnarray}
Since the new longitudinal boundary condition (\ref{NBC}) is now synchronous and periodic, one can use an expansion in longitudinal Fourier modes. However, under the MFL conditions only the longitudinal mode closest to $\omega$ has components different from zero. This mode corresponds to a zero longitudinal frequency so that $\partial_{z'} F = 0$. 
Finally, we note that $\Gamma$ at zeroth order is one so that $F\approx E$, we divide the equation by $T/2$, consider $L=O(T/2)$ and introduce the renormalised parameters:
\begin{eqnarray}
\label{NPAR2_app}
\theta = \frac{\Theta} {T/2}; \;\;\;\;\;\;\;\;
a = \frac {\Lambda-L} {kT}; \;\;\;\;\;\;\;\; \kappa = \frac{T}{2\tau}
\end{eqnarray}
to obtain:
\begin{eqnarray}
\partial_{\kappa t'} E = E_{I} -(1+i\theta) E + i \eta |E|^2 E + i a \nabla^2 E . 
\label{LLE_app}
\end{eqnarray}
where $E_I$ has been normalised by $\sqrt{T/2}$. Eq. (\ref{LLE_app}) is the renowned spatial Lugiato-Lefever LLE model \cite{Lugiato1987}.

We note that in the transverse case above we considered a single longitudinal mode of the cavity. When deriving the LLE in the time domain, instead, one considers a single transverse mode but a very large number of longitudinal cavity modes. Each longitudinal mode is considered to be slowly varying and with a cavity linewidth much smaller than the free spectral range. In this case a temporal LLE (\ref{FTLLE}) can be obtained under appropriate MFL conditions \cite{Castelli2017}.

\end{document}